\documentclass[aps,prl,twocolumn,superscriptaddress,nofootinbib,floatfix]{revtex4-2}
\usepackage{amsmath,amssymb}
\usepackage{graphicx}
\usepackage{booktabs}
\usepackage{dcolumn}
\usepackage{float}
\usepackage{xcolor}
\usepackage{hyperref}
\definecolor{reflink}{rgb}{0.00,0.27,0.78}
\definecolor{refcite}{rgb}{0.72,0.10,0.10}
\definecolor{refurl}{rgb}{0.00,0.44,0.44}
\hypersetup{
  colorlinks=true,
  linkcolor=reflink,
  citecolor=refcite,
  urlcolor=refurl,
  filecolor=refurl,
  linktoc=all,
  pdfborder={0 0 0},
  bookmarksnumbered=true,
  pdfauthor={Bo-Qiang Lu},
  pdftitle={Cosmic-string calorimetry with a 220 PeV neutrino},
}
\makeatletter
\let\plain@eqref\eqref
\renewcommand\eqref[1]{\textcolor{reflink}{\plain@eqref{#1}}}
\makeatother

\begin{document}
\setlength{\textfloatsep}{8pt plus 2pt minus 2pt}
\setlength{\intextsep}{6pt plus 2pt minus 2pt}
\setlength{\floatsep}{8pt plus 2pt minus 2pt}

\title{Cosmic-string calorimetry with a 220\,PeV neutrino}

\author{Bo-Qiang~Lu}
\affiliation{School of Science, Huzhou Normal University, Huzhou, Zhejiang 313000, P.~R.~China}

\begin{abstract}
KM3NeT has reported KM3-230213A, a $220$\,PeV neutrino with no source identified.
If it comes from a cosmic-string network, energy conservation makes the event a
calorimeter. The flux measures the power the network has spent, not how it emits,
and a cusp at critical current can supply that power at unit efficiency. The event
gives $G\mu\epsilon_p=(2.8\text{--}4.3)\times10^{-11}$ for $f_\nu=0.1$. Pulsar
timing already reaches $10^{-10}$, so one event competes with the arrays. For the
branch that still radiates gravitationally, flux and timing together require a
third of the network's energy in particles, a share that further events above
$100$\,PeV would measure.
\end{abstract}

\maketitle

\begingroup\renewcommand\thefootnote{}\footnotetext{%
\textsuperscript{*}Electronic address: \href{mailto:bqlu@huznu.edu.cn}{\textcolor{black}{bqlu@huznu.edu.cn}}}%
\addtocounter{footnote}{-1}\endgroup

\section*{Introduction}

Cosmic neutrinos carry no charge, and magnetic fields do not deflect them, so an
event above $10^{17}$~eV should point back to its source. The first such event,
KM3-230213A, is a through-going muon of reconstructed energy
$220^{+570}_{-110}$~PeV recorded by the KM3NeT Collaboration \cite{KM3NeT}. The flux,
$E^2\Phi_\nu\simeq5.8\times10^{-8}$~GeV\,cm$^{-2}$\,s$^{-1}$\,sr$^{-1}$ per flavor, 
exceeds the IceCube limit at $10^{18}$~eV \cite{ICehe} by a factor of
seventeen. The excess is a $3.5\sigma$ tension for a single power law
\cite{LiMT}, $2.9$--$3.1\sigma$ in analyses restricted to the extremely-high-energy (EHE) data 
or built on a transport equation \cite{ICehe,PRT}. Fifteen years of IceCube
data in the direction of the event show no steady or flaring counterpart
\cite{IPSO}. Over the combined range from $10^{16}$ to $10^{26}$~eV, little room
is left for a second event of that kind \cite{Troitsky}. Energy conservation makes the
event a calorimeter for the network.

The proposals differ in where the new physics sits. At the source, several groups invoke a
dark-sector decay, a superheavy relic \cite{Kohri,Borah,Jho}, vector dark matter
\cite{Su}, or a sterile neutrino from an evaporating black hole \cite{Choi},
with the abundance set at a first-order transition that also leaves a
gravitational-wave background \cite{Cline}. Primordial black holes
enter as a source that the event bounds \cite{Boccia,Klipfel,Airoldi}, the
memory-burden effect keeping them viable \cite{Zantedeschi}. In propagation,
neutrino self-interactions \cite{He}, scattering on dark matter
\cite{Bertolez}, dark radiation from cosmic-ray collisions \cite{Narita}, and a
sterile component that converts on the way to the detector \cite{Brdar}
have been considered. The event itself has been read as a bound on Lorentz
violation \cite{KM3LIV,Satunin,Yang}, or as a normalization of a cosmogenic or
transient source
\cite{KM3cosmo,Neronov,Fang}, fitted in each case with one free parameter.

In this paper we take a cosmic-string network as the source and ask what the
event demands of it. Among those proposals a string appears only as a source of gravitational waves in
a dark-sector model \cite{Su,Khan}, never as the source of a single event. Of the sixty-two papers the event has drawn so far,
sixteen mention a string, and none takes one as the source. The one fit
to the diffuse flux absorbs the
tension into a normalization, returning no $G\mu$ \cite{CHK23}. A string is
the line-like defect left when a $U(1)$ symmetry breaks spontaneously, about one
per horizon volume \cite{Kibble}. In the Abelian Higgs model it is the
Nielsen--Olesen vortex, a flux tube of width $1/\eta$ and tension
$\mu\sim\eta^2$, the mass per unit length, whose gravitational phenomenology depends on the single
combination $G\mu\sim(\eta/m_{\rm Pl})^2$ \cite{HK}. A grand-unified transition
gives $G\mu\sim10^{-6}$. A late $U(1)$, a dark sector, or the warped throat of
brane inflation, where F- and D-strings survive as cosmic superstrings
\cite{CMPSt}, place it far lower. Pulsar timing bounds the gravitational-wave
amplitude at the level of $10^{-10}$ \cite{NANOGrav,EPTA,PPTAstring}. Ground-based
detectors reach $G\mu\lesssim4\times10^{-15}$ in their band, but only for the
one-loop distribution they assume, and the other distributions weaken that bound
\cite{LVK}. A network that supplies the event falls where timing has begun to probe.

A particle channel appears once the string carries a current, as a
superconducting string \cite{Witten} does. Charge carriers condense into a current $I=i_c\eta$, with $i_c\le1$, which
carries $i_c^2\mu$ per unit length and makes the string a source of particles
as well as gravitons. Superconducting strings were proposed early as
sources of ultrahigh-energy neutrinos \cite{HSW87,Plaga}. The fluxes quoted since
span orders of magnitude, from the discharge of the current at a cusp
\cite{BO09}, through the emission of moduli there \cite{BSV11}, to the ejection
of the charge carriers themselves \cite{HSW87,Vachaspati}. 

In the scaling regime a network processes energy at a rate fixed by its tension and the expansion rate,
whatever the distribution of its loops in length. If a fraction
$\epsilon_p$ of that power reaches particles, the flux at $E$ follows from
energy conservation alone. Inverted, the
relation measures the power the network has spent,
with the loop production function and the field filling factor absorbed into
$\epsilon_p$. The cascade that populates the spectrum also fixes its width, 
so the string scale enters only through the cut-off and one free parameter remains.

Earlier string calculations start from the opposite end
\cite{BO09,Lunardini,ASV}. They integrate a loop production function over length and
time, weight each loop by a burst rate, a beaming factor and an ambient
field, and sum the bursts. The flux then carries several powers of $G\mu$
and depends on quantities that are modeled rather than measured. In this approach the loop distribution drops out of the rate and fixes only the
maximum energy and the shape of the spectrum (Sec.~S1). The efficiency $\epsilon_p$ is not calculated from first
principles, but the event measures how the network power is partitioned, and a
second event would fix the split. For KM3-230213A we find
$G\mu\epsilon_p=(2.8\text{--}4.3)\times10^{-11}$ with $\epsilon_p\gtrsim0.35$.

\section*{Energy budget of a scaling network}

In the scaling regime the long-string energy density is
\begin{equation}
\rho_\infty=\zeta\,\mu H^2,
\label{eq:rho}
\end{equation}
where $\zeta=1/(4\gamma^2)$ in the matter era and $\gamma=d/d_h=0.17$ is the
ratio of the mean string separation to the horizon $d_h=3t=2/H$ \cite{BOS}.
Simulations of the loop production function give $\zeta=8.7$ in this era and
$\zeta=44$ deep in the radiation era. The flux is dominated by the matter era,
and we use the first value throughout (Appendix). The long-string component
obeys $\dot\rho_\infty+2H\rho_\infty=-Q$, and $\rho_\infty\propto t^{-2}$ in
scaling then fixes the injection rate,
\begin{equation}
Q=\kappa\,\rho_\infty H,\qquad \kappa=1,
\label{eq:gamma}
\end{equation}
with $\kappa=2$ in the radiation era. Nothing in Eq.~\eqref{eq:gamma} refers to
the loop distribution. Both coefficients drift as the Universe leaves matter domination, by about
$40\%$ at $z\lesssim0.3$, and that drift limits the accuracy of the tensions
quoted here.

The loops decay by gravitational radiation \cite{DV} and, where a particle
channel operates, by particle emission. Let $\epsilon_p\le1$ be the fraction of
$Q$ that reaches particles. The cascade spreads that power uniformly in $\ln E$
below the cut-off. The complementary power leaves as gravitational waves, 
and pulsar timing bounds it directly. Of the particle energy, a fraction $f_\nu$ emerges as neutrinos of one flavor in a
logarithmic interval $\Delta\ln E$ around $E$. The emissivity per unit $\ln E$ is
then
\begin{equation}
q_{\ln E}=\zeta\,\frac{8\pi}{3}\,\kappa\,G\mu\,\rho_c H\;
\frac{\epsilon_p f_\nu}{\Delta\ln E},
\label{eq:emissivity}
\end{equation}
where the factor $8\pi/3$ converts $\mu H^2$ into a fraction of the critical
density. Integrating over redshift for isotropic emission,
\begin{equation}
\begin{split}
E^2\Phi_\nu&=\frac{2}{3}\,\kappa\zeta\,\frac{c}{H_0}\,\rho_c H_0\,R\,
G\mu\,\frac{\epsilon_p f_\nu}{\Delta\ln E}\\
&\simeq4.3\times10^{5}\,G\mu\,\frac{\epsilon_p f_\nu}{\Delta\ln E},
\end{split}
\label{eq:master}
\end{equation}
in GeV\,cm$^{-2}$\,s$^{-1}$\,sr$^{-1}$, where $R=0.47$ for $\Omega_m=0.31$ and
$\Omega_\Lambda=0.69$ integrated to $z=30$.

Equation~\eqref{eq:master} is the string analogue of the calorimetric bound of
Waxman and Bahcall \cite{WB}, with the scaling energy density in place of the
observed luminosity density. Since the flux measures the power the network has
spent rather than the details of its emission, one event reads out the total energy
the network has released.

\section*{The efficiency the event requires}

Setting Eq.~\eqref{eq:master} equal to the flux of KM3-230213A fixes
$\epsilon_p$ at each tension. We take the event as one draw from an isotropic population 
with the observed rate. The Poisson range on the count, $0.1<N<3.9$ with each
bound at $90\%$ confidence, then covers that rate.
A transient origin leaves Eq.~\eqref{eq:gmu} intact, since the fluence of a single
burst fixes the energy the network spent on it, and removes only the counting
interval.

A cascade started by a particle at the string scale runs from the QCD scale to
$E_{\max}=i_c\eta$ with equal energy per decade, and the width is then not a free
parameter,
\begin{equation}
\Delta\ln E=\ln\frac{E_{\max}}{1~{\rm GeV}}
\simeq\ln\frac{\sqrt{G\mu}\,m_{\rm Pl}}{1~{\rm GeV}},
\label{eq:width}
\end{equation}
with the tension fixed by the flux. Solving
Eqs.~\eqref{eq:master} and \eqref{eq:width} together gives
\begin{equation}
G\mu=4.3\times10^{-11},\qquad \eta=8.0\times10^{13}~{\rm GeV},
\label{eq:fixed}
\end{equation}
with $\Delta\ln E=32$. The cascade then reaches $8\times10^{22}$~eV in particles, its neutrinos a factor of twenty lower. A charged condensate would discharge into an electromagnetic cascade and
leave no neutrinos, so the carriers are taken to carry color and the cascade to be
hadronic from the start. A hadronic cascade leaves
at most about a tenth of its energy in one neutrino flavor \cite{Kelner}, so we take
$f_\nu=0.1$ and obtain
\begin{equation}
G\mu\,\epsilon_p\simeq4.3\times10^{-11}.
\label{eq:gmu}
\end{equation}
Widening the count to the full Poisson range multiplies this by forty, and
$G\mu$ runs from
$4.3\times10^{-12}$ to $1.7\times10^{-10}$ at $\epsilon_p=1$, each end moving by
three percent once the width is re-solved.

Pulsar timing bounds the gravitational-wave amplitude, which scales as
$G\mu(1-\epsilon_p)$ because the particle channel takes the complementary share.
The NANOGrav 15~yr limit runs from $7.9\times10^{-11}$ to $2.1\times10^{-10}$
across the four loop distributions it fits, and EPTA gives $2\times10^{-10}$
\cite{NANOGrav,EPTA}. We adopt the smallest, which gives the strongest
requirement on $\epsilon_p$. Over the other three the floor falls to $0.24$,
$0.18$ and $0.17$. The
scaling with $(1-\epsilon_p)$ holds at fixed loop distribution. Particle emission shortens the
life of a loop and raises the number density, but the two effects largely cancel
at fixed frequency, so the main imprint is a cut-off at high frequency
\cite{ASV} (Sec.~S2 of the Supplemental Material \cite{SM}). Combining this bound
with Eq.~\eqref{eq:gmu},
\begin{equation}
\epsilon_p\gtrsim\frac{(G\mu)_0}{(G\mu)_0+B},\qquad
(G\mu)_0=4.3\times10^{-11}\,\frac{0.1}{f_\nu},
\label{eq:eps}
\end{equation}
with $B=7.9\times10^{-11}$. For the central count the floor is $\epsilon_p\gtrsim0.35$,
about a third, since the tension the flux requires is within a factor of two of
what timing allows. Vector radiation takes a share $\epsilon_V$ of the budget and lowers the floor to
$(1-\epsilon_V)(G\mu)_0/[(G\mu)_0+B]$, which is $0.28$ at $\epsilon_V=0.2$. The Poisson range of
the count carries $\epsilon_p$ from $0.05$ to $0.68$, and over $f_\nu=0.05$--$0.3$ it moves
from $0.52$ to $0.15$ (Appendix). 
A currentless string obeys this bound. A current suppresses the gravitational
channel and, in the fields where the cusp is bright, removes it (Appendix). Equations~\eqref{eq:width} and \eqref{eq:fixed} leave nothing free. The same
cascade fixes the flux at $100$~TeV, where that flux exceeds the measured diffuse
flux by a factor of four, and at $1$~EeV, where the EHE limit places the top of the
spectrum. Photons from the same cascade pair-produce on the cosmic microwave
background and reach the GeV band, where a steady population overshoots the extragalactic
gamma-ray background by an order of magnitude, while a single burst is exempt. The
dark ages, which bound the power deposited in the gas, give
$G\mu\lesssim7\times10^{-11}$ (Appendix).

\begin{figure}[t]
\includegraphics[width=\columnwidth]{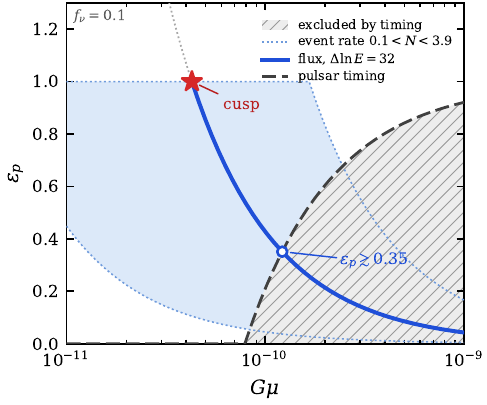}
\caption{Particle efficiency of a string network. The solid curve is the
efficiency the flux of KM3-230213A requires at $f_\nu=0.1$, dotted in gray where
it exceeds unity, and the pale curves bracket the count $0.1<N<3.9$. Timing
excludes the hatched region, the pale band survives both, and the open point marks
their meeting. The star is the critical cusp.}
\label{fig:wedge}
\end{figure}

Figure~\ref{fig:wedge} combines the two constraints. The flux requires the
network to lie on the falling curve, timing excludes everything below the rising
one, and a solution exists only where the curve stays above the bound. The two
meet at $G\mu=1.2\times10^{-10}$, above the bound that timing sets at vanishing
efficiency. There the required efficiency is smallest, and the two observations
bound the particle and gravitational shares from opposite sides. The event adds
the partition, not a tighter bound on the tension.

\section*{The current-carrying cusp}

At a cusp, where two segments of the string meet at relativistic speed, the
current $I=i_c\eta$ is discharged into a spectrum
$dN/dE\propto E^{-2}$ up to $E_{\max}=i_c\eta$
\cite{BO09,Lunardini,Vachaspati}. Beaming does not by itself suppress the diffuse flux. The cone reduces the number of
bursts that point at us by $4\pi/\Omega$ and raises their fluence by the same
factor, so the two cancel.
The share of the loop energy that the current can carry away is set by the ratio
of the two luminosities,
\begin{equation}
r\simeq50\,f_{-3}B_{-6}i_c\,\frac{1}{\eta_{10}}\frac{\ell}{\ell_{\min}},
\qquad \ell_{\min}=\Gamma_{\rm GW}G\mu t,
\label{eq:ratio}
\end{equation}
with $f_{-3}=f_B/10^{-3}$, $B_{-6}$ the ambient field in $\mu$G,
$\eta_{10}=\eta/10^{10}$~GeV, and
$\Gamma_{\rm GW}\simeq50$ \cite{BO09}. Loops form with $\ell\simeq\alpha t$ and
$\alpha\simeq0.1$ \cite{BOS}, well above $\ell_{\min}$, and with
$\eta=\sqrt{G\mu}\,m_{\rm Pl}$, Eq.~\eqref{eq:ratio} becomes
\begin{equation}
r\simeq8\times10^{-11}\,\frac{i_cB_{-6}f_{-3}}{(G\mu)^{3/2}}.
\label{eq:ratio2}
\end{equation}
At $G\mu=4.3\times10^{-11}$ this is $3\times10^{5}$, and the cusp supplies the
event at unit efficiency, with the tension and $\eta$ of Eq.~\eqref{eq:fixed}.

\begin{figure}[t]
\includegraphics[width=\columnwidth]{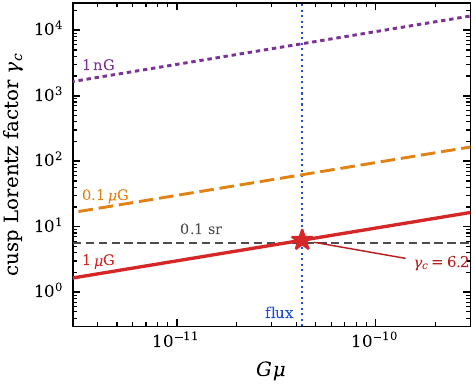}
\caption{Cusp Lorentz factor from Eq.~\eqref{eq:gammac} at the tension the flux
requires. It scales as $\sqrt{G\mu}$ and inversely with the field. The vertical
line is $G\mu=4.3\times10^{-11}$, where the event sits, and the star marks
$\gamma_c=6.2$ for a microgauss field. The horizontal line is $\gamma_c=5.6$,
below which the cone exceeds $0.1$~sr. A $1$~nG field raises it by three orders and
makes a pencil beam, the diffuse flux unchanged.}
\label{fig:cusp}
\end{figure}

There the cusp is critical. The Lorentz factor follows from the
ratio of the critical current to the current the ambient field induces, and at the
formation length $\ell=\alpha t$ it is
\begin{equation}
\gamma_c\simeq9.5\times10^{5}\,\frac{i_c\sqrt{G\mu}}{B_{-6}},
\label{eq:gammac}
\end{equation}
which is $6.2$ for microgauss fields at the tension of the event
(Fig.~\ref{fig:cusp}). The length that enters is the formation length. Evaluated at the minimum length,
shorter by $5\times10^{7}$, the same equation gives $3\times10^{8}$ and
reverses the scaling in $G\mu$ (Sec.~S2). Saturation follows from $\gamma_c=1$ at
$\ell\simeq6\alpha t$, beyond which the induced current would exceed the critical
one. A loop at or above the formation length therefore carries its
current at the critical value.

Field and tension enter Eqs.~\eqref{eq:ratio2} and \eqref{eq:gammac} in opposite directions, so the beaming depends on the environment. In the fields
where the channel is bright, $B\simeq1\,\mu$G with $f_B\sim10^{-3}$, the cone
$\pi/\gamma_c^2$ is $0.08$~sr, a burst that points at us is brighter by
$4\pi/\Omega\simeq1.5\times10^{2}$, and the tension the flux requires puts the cusp
just inside the $0.1$~sr line of Fig.~\ref{fig:cusp}. In the intergalactic field,
$B\simeq1$~nG, Eq.~\eqref{eq:gammac} gives $\gamma_c\simeq6\times10^{3}$ and the
cone narrows to $8\times10^{-8}$~sr. Both environments reproduce the event.

A cusp bright enough to supply the event reaches a tip Lorentz factor of a few in the 
fields where the channel is defined, its spectrum cuts off at $i_c\eta\simeq\eta$, and it releases the whole loop energy. The current that
produces this radiation also closes the gravitational channel, since the emission
efficiency of a cusp falls as $(1-i_c)^2$ \cite{RS22}. At $i_c\simeq1$ the
cusp channel is silent and the timing bound does not apply to it (Appendix). If the
current is not replenished after discharge, the released fraction is $i_c^2$ and
the tension rises to $4.3\times10^{-11}/i_c^2$. The channel is then excluded for
$i_c\lesssim0.7$, once the gravitational-wave signal is counted against that
bound.

The cut-off is where the EHE limit enters. At critical current $E_{\max}=i_c\eta$
is $8\times10^{22}$~eV, and an $E^{-2}$ spectrum normalized to the flux of the event
reaches $1$~EeV at seventeen times the limit. A cut-off at $1$~EeV takes $i_c\simeq1.4\times10^{-5}$, and
Eq.~\eqref{eq:ratio2} stays above unity down to a few times $10^{-6}$, so the
discharge keeps its efficiency as it loses its beaming. Either way the diffuse flux is the same, and the two corners differ only in burst rate and
fluence (Sec.~S3).

The burst-by-burst calculation of Ref.~\cite{BO09} also rests on energy conservation,
but its rate is set by the loop production function, its current by the ambient
field, and its fluence by the beaming. The resulting flux carries no factor of $\gamma_c$
and keeps its value as $G\mu$ falls, while Eq.~\eqref{eq:master} falls in proportion.
The two agree at $G\mu\simeq1.6\times10^{-11}$ and separate below it
(Sec.~S4).

\section*{Other channels}

Cusp annihilation in the Hill--Schramm--Walker picture \cite{HSW87,VHS} releases
the bulk energy $\mu\ell$ with a fragmentation spectrum close to $E^{-2}$ below
its peak \cite{Bhattacharjee}, and $\epsilon_p$ can be of order unity. No other channel in the catalog of Sec.~S5 reaches order-unity efficiency
without a separate argument, and the parameter that makes this one efficient also
suppresses its gravitational-wave signal. A loop oscillates
$\sim1/(\Gamma_{\rm GW}G\mu)$ times and a cusp releases a fraction $\delta$ of
the loop energy, so $\epsilon_p\sim\min(1,\delta/(\Gamma_{\rm GW}G\mu))$, which
saturates for any $\delta\gtrsim2\times10^{-9}$. With $\epsilon_p=1$ the channel
sits at
$G\mu\simeq4.3\times10^{-11}$, and the floor of $0.35$ applies to it, leaving
two thirds of its power in gravitational waves.

A fireball \cite{Plaga,Babul,BHV,GruzinovVilenkin} cools as a quasi-thermal
plasma at $T_f\simeq1$~GeV for $\eta=10^{14}$~GeV in a $1$~nG field, and the
cusp boost lifts its photons to $8$~TeV \cite{GruzinovVilenkin}, so the flux at
$220$~PeV is suppressed by $\exp(-2.8\times10^{4})$. Vortons \cite{Carter,MartinsShellard,Lemperiere}, necklaces \cite{BV97} and
black holes from collapsing cusps \cite{PZ} store or inject their energy at formation, 
and an abundance, a lifetime or a bead mass replaces $G\mu$. Each can be tuned to the flux, at
the cost of the prediction, because the required mass or abundance must
fall inside a window thirty-five orders of magnitude wide (Sec.~S5).

\section*{Conclusions}

Read as a calorimeter, the event converts the energy budget of a scaling network
into its tension and the share of its power that reaches particles. The neutrino
flux measures the particle channel, the timing residual the gravitational one, and
the EHE limit the top of the spectrum. Together they give
$G\mu\epsilon_p=(2.8\text{--}4.3)\times10^{-11}$, the range set by the cut-off,
at the string scale for a critical current and at $1$~EeV for a
subcritical one, with $\epsilon_p\gtrsim0.35$ at the tightest of the four
loop distributions. The current-carrying cusp exhausts the
flux at the upper end, with its tip at $\gamma_c\simeq6.2$ and its beaming set by
the environment rather than by the network. Bulk ejection, less calculable but
needing no such argument, is the minimal realization of the budget.

For any channel that still radiates gravitationally, the background scales as
$(1-\epsilon_p)/\epsilon_p$, from $7.9\times10^{-11}$ at the timing bound to
$4\times10^{-13}$ at $\epsilon_p=0.99$, with particle emission cutting it off at
high frequency \cite{OkadaSeto}. A detection anywhere in that range, with the
flux, measures the share of power spent on particles, and a bound at $10^{-12}$
would put it above $0.98$, where the Hill--Schramm--Walker picture sits. The cusp
lies outside that range, its current suppressing the gravitational radiation, so a
detection would point to annihilation.

Every event above $100$~PeV adds a point
to Fig.~\ref{fig:wedge} and narrows $\epsilon_p$ as the square root of the
count. The completed KM3NeT
array, with $230$ detection units against the $21$ that recorded this event
\cite{KM3NeT}, would collect of order five events a year at this flux. Thirty
events, six years of the completed array, would pin $\epsilon_p$ to $\pm0.06$
statistically (Sec.~S3), against a systematic error set by $f_\nu$, and PUEO would
reach the $1$~EeV cut-off \cite{PUEO}. Energy conservation is the least model-dependent
statement about a string network, and for this event it is also the
most restrictive.

\section*{Acknowledgments}
BQL is supported in part by the National Natural Science Foundation of China under Grant No. 12405058.

\section*{Appendix}

\subsection*{Derivation of Eq.~\eqref{eq:master}}
\label{em:derivation}

In a radiation-dominated universe $H=1/(2t)$ and Eq.~\eqref{eq:gamma} gives
$\kappa=2$. The flux is dominated by $z\lesssim5$, where $\kappa=1$ applies. Deep in the radiation era the rise to two affects only a tail suppressed
by $(1+z)^{-2}$. The factor $8\pi/3$ in Eq.~\eqref{eq:emissivity}
follows from $\mu H^2=(8\pi/3)G\mu\,\rho_c$. For isotropic emission the flux per
unit solid angle from a shell of thickness $dr$ is $q_{\ln E}dr/4\pi$, and two
powers of $1+z$ convert the emitted energy and its bin to observed values while
three convert the proper emissivity into a comoving one, which gives
\begin{equation}
E^2\Phi_\nu=\frac{c}{4\pi}\frac{\kappa\zeta\mu H_0^2}{\Delta\ln E}\,
\epsilon_p f_\nu\int_0^{z_{\max}}\frac{E(z)^2\,dz}{(1+z)^5},
\end{equation}
with $E(z)\equiv\frac{H(z)}{H_0}$.
For $\Omega_m=0.31$ the integral is $R=0.4725$ at $z_{\max}=30$, and in
Einstein--de Sitter the same integral is $1-1/(1+z_{\max})$, which is $0.968$ at
that $z_{\max}$. The numerical coefficient of
Eq.~\eqref{eq:master} follows with $\rho_c=5.3\times10^{-6}$~GeV\,cm$^{-3}$,
$c/H_0=1.3\times10^{28}$~cm and $H_0=2.3\times10^{-18}$~s$^{-1}$. The two powers
of $H_0$ cancel, and the grouping is kept because $\mu H_0^2$ is what
Eq.~\eqref{eq:gamma} supplies.

Since every tension scales as $1/(\kappa\zeta)$, the two conventions for the
long-string density have to be kept apart. Simulations quote
$\rho_\infty=\mu/(\gamma d_h)^2$ with $\gamma=d/d_h$ \cite{BOS}. In the matter era
$d_h=3t=2/H$ and $\gamma=0.17$, hence $\zeta=1/(4\gamma^2)=8.7$. In the radiation
era $d_h=2t=1/H$ and $\gamma=0.15$, hence $\zeta=1/\gamma^2=44$. The largest uncertainties are therefore in $\gamma$, which enters
$\zeta$ as $1/\gamma^2$, and in the drift of $\kappa$, both of which affect every
channel alike.

\subsection*{The width of the spectrum}
\label{em:sens}

Equation~\eqref{eq:width} states that a network has one energy scale, since the
cascade of a particle of energy $\eta$ populates every decade below it with equal
energy. The width and the tension are then one variable, and
Eq.~\eqref{eq:fixed} is the fixed point of the pair. The iteration converges
because Eq.~\eqref{eq:gmu} depends on $G\mu$ only logarithmically. From
$\Delta\ln E=30$ one pass gives $G\mu=4.01\times10^{-11}$ and $\Delta\ln E=32.00$,
a second gives $4.28\times10^{-11}$ and $32.01$. The cascade is
an assumption and not a derivation. A hadronic cascade puts the neutrino cut-off a
factor of twenty below $\eta$, which lowers $\Delta\ln E$ by three and every
tension by ten percent.

Emission confined to a decade would need a cut-off at $2\times10^{9}$~GeV, hence
$i_c\simeq10^{-4}$ and a tension lower by a factor of fourteen. The network does
not forbid that corner, but a calculation that assumes a cascade cannot choose it.

The formal bound with $\epsilon_pf_\nu\le1$ and $\Delta\ln E\ge1$ is
$G\mu>1.3\times10^{-13}$, independent of the channel. For thirty-two e-folds it
rises to $4.3\times10^{-12}$, and with $f_\nu=0.1$ to $4.3\times10^{-11}$. The
value of $f_\nu$ carries the largest astrophysical uncertainty, entering every
tension inversely and the bound of Eq.~\eqref{eq:eps} directly. At
$f_\nu=0.05$ the bound rises from $0.35$ to $0.52$. Table~\ref{tab:sens}
collects, over the range of $f_\nu$, the tension at unit efficiency, the floor it
implies, and the Lorentz factor of the cusp at that tension.

\begin{table}[H]
\caption{The tension at $\epsilon_p=1$, the floor from Eq.~\eqref{eq:eps} with
$B=7.9\times10^{-11}$, and the cusp Lorentz factor at that tension in a microgauss
field, over the flavor fraction, the only input left once Eq.~\eqref{eq:width}
fixes the width. The count scales $(G\mu)_0$ by $N$ and $\gamma_c$ by $\sqrt N$,
and $\gamma_c$ stays above unity, so the cusp reaches its critical current while
still beamed.}
\label{tab:sens}
\footnotesize
\setlength{\tabcolsep}{4.0pt}
\begin{ruledtabular}
\begin{tabular}{cccc}
$f_\nu$ & $(G\mu)_0$ & $\epsilon_p^{\min}$ & $\gamma_c$ \\
\hline
$0.05$ & $8.6\times10^{-11}$ & $0.52$ & $8.8$ \\
$0.10$ & $4.3\times10^{-11}$ & $0.35$ & $6.2$ \\
$0.20$ & $2.1\times10^{-11}$ & $0.21$ & $4.4$ \\
$0.30$ & $1.4\times10^{-11}$ & $0.15$ & $3.6$ \\
\end{tabular}
\end{ruledtabular}
\end{table}

\subsection*{The current and the gravitational channel}
\label{em:current}

The bound of Eq.~\eqref{eq:eps} applies to a currentless string, because a
current suppresses the gravitational emission of cusps and kinks. In the
transonic model the efficiency follows
\begin{equation}
\Gamma_{\rm GW}=\Gamma_{\rm NG}\,(1-|F'|)^{B_\Gamma},\qquad B_\Gamma\simeq2,
\label{eq:rs}
\end{equation}
for the symmetric currents in which cusps exist at all \cite{RS22}. The exponent
is $1.5$ for kinks and smaller by a factor $\sqrt2$ for chiral currents, which
have no cusps. Since the current carries an energy $i_c^2$ per unit length in units of
$\mu$, the $i_c$ of Eq.~\eqref{eq:ratio} is $|F'|$, and the timing bound becomes
$G\mu(1-\epsilon_p)(1-i_c)^2<B$. The floor of Eq.~\eqref{eq:eps} generalizes to
\begin{equation}
\epsilon_p\gtrsim\frac{(G\mu)_0}{(G\mu)_0+B\,(1-i_c)^{-2}},
\label{eq:epsic}
\end{equation}
which gives $0.31$ at $i_c=0.1$, $0.26$ at $0.2$, $0.12$ at $0.5$, and no bound at
all at the critical current, where the string is tensionless and does not radiate.

The ratio of Eq.~\eqref{eq:ratio} exceeds unity, and the cusp supplies the event
at unit efficiency, whenever $i_cB_{-6}f_{-3}\gtrsim3\times10^{-6}$. The condition
fails only if the network power is carried by loops within a factor of $150$ of $\ell_{\min}$, rather than at the formation length $\ell\simeq\alpha
t$ assumed here.

The cusp sits at that end of the sequence, since Eq.~\eqref{eq:gammac} reaches
$\gamma_c\simeq6$ in a microgauss field only if $i_c\simeq B_{-6}$, and the
current that makes the cusp wide is then the current that silences it. There Eq.~\eqref{eq:eps} imposes no bound, and the event requires no
gravitational-wave counterpart. Vector radiation takes a
share of the remaining budget \cite{RS24}, so the partition is three-way and the
gravitational share no larger. The floor of Eq.~\eqref{eq:eps}
therefore applies to the channels that still radiate gravitationally, and the data
select cusp annihilation among them.

\subsection*{The diffuse and gamma-ray backgrounds}
\label{em:data}

Whatever the mechanism, an isotropic population that reproduces
$5.8\times10^{-8}$ at $220$~PeV also reproduces it at $100$~TeV, where the IceCube
flux is $1.44\times10^{-8}$ \cite{NST}. An $E^{-2}$ spectrum therefore
over-produces the diffuse flux by a factor of about four, the $3.5\sigma$ tension
that Ref.~\cite{LiMT} finds for any isotropic single power law
\cite{HESE,Troitsky}. The tension lies in the data rather than in the model,
since the normalization is the measurement. The EHE limit constrains the other end of the same spectrum.
Equation~\eqref{eq:width} puts the top at $8\times10^{22}$~eV, seventeen times the
limit at $1$~EeV, and a cut-off at $1$~EeV removes the excess. A transient origin
does not supply one, since each decade of an $E^{-2}$ burst carries the same
energy. A subcritical current does, because $E_{\max}=i_c\eta$ falls below
$1$~EeV for $i_c\lesssim10^{-5}$. The two cut-offs bracket the budget. With $\Delta\ln E$ free, Eq.~\eqref{eq:master} fixes $G\mu\epsilon_p/\Delta\ln E$ alone,
so a cut-off at $1$~EeV lowers the width to $21$ and the tension to
$2.8\times10^{-11}$. The bound holds at either cut-off, between $2.8$ and
$4.3\times10^{-11}$. Table~\ref{tab:corner} follows the sequence.

\begin{table}[H]
\caption{The corner in which the cut-off sits at $E_{\rm br}$.
$i_c=E_{\rm br}/\eta$ is the current it requires, $\gamma_c$ and $r$ follow from
Eqs.~\eqref{eq:gammac} and \eqref{eq:ratio2}, and $\epsilon_p=r/(1+r)$,
$\Delta\ln E$ and $G\mu$ are solved with Eq.~\eqref{eq:master} at $f_\nu=0.1$ in a
microgauss field. $\gamma_c<1$ marks an isotropic discharge.}
\label{tab:corner}
\footnotesize
\begin{ruledtabular}
\begin{tabular}{cccccc}
$E_{\rm br}$ [GeV] & $i_c$ & $\gamma_c$ & $\epsilon_p$ & $\Delta\ln E$ & $G\mu$ \\
\hline
$10^{9}$ & $1.4\times10^{-5}$ & $7.8\times10^{-5}$ & $0.87$ & $20.7$ & $3.2\times10^{-11}$ \\
$10^{10}$ & $1.5\times10^{-4}$ & $7.8\times10^{-4}$ & $0.99$ & $23.0$ & $3.1\times10^{-11}$ \\
$10^{11}$ & $1.4\times10^{-3}$ & $7.8\times10^{-3}$ & $1.00$ & $25.3$ & $3.4\times10^{-11}$ \\
$10^{12}$ & $1.4\times10^{-2}$ & $7.8\times10^{-2}$ & $1.00$ & $27.6$ & $3.7\times10^{-11}$ \\
$10^{13}$ & $0.13$ & $0.78$ & $1.00$ & $29.9$ & $4.0\times10^{-11}$ \\
$8\times10^{13}$ & $1.00$ & $6.2$ & $1.00$ & $32.0$ & $4.3\times10^{-11}$ \\
\end{tabular}
\end{ruledtabular}
\end{table}

In a hadronic cascade each neutral pion gives two photons of comparable energy, and the model predicts a $100$~PeV photon flux of the same order as the neutrino
flux. The photons pair-produce on the cosmic microwave background and cascade
into the GeV band, where the extragalactic background bounds the total energy
injected \cite{Fang}. For a population at $5.8\times10^{-8}$ the cascade flux
exceeds the observed background by an order of magnitude \cite{Fang}. The
constraint follows from the population and disappears for a single burst, which
contributes negligibly to the integrated background. 
The dark ages bound any channel that deposits its power in the gas, excluding
vector-radiating networks above $G\mu\simeq10^{-13}$ when the whole injected
energy reaches the gas, because the deposited power erases the 21~cm absorption at $z\simeq89$
\cite{Afzal,Si}. The hadronic cascade deposits far less. At $z=90$ the pairs cool
by inverse Compton on the CMB before reaching the keV window where the gas
absorbs, so $3\times10^{-3}$ of the electromagnetic energy reaches the gas. Half
of the particle energy starts as electromagnetic, so $1.5\times10^{-3}$ of the
total power is deposited, and rescaling by its inverse turns $10^{-13}$ into
$G\mu\lesssim7\times10^{-11}$, an order-of-magnitude conversion that lands within a
factor of two of the flux's own tension.

\clearpage
\onecolumngrid

\begin{center}
{\large\bfseries Supplemental Material}\\[3pt]
{\bfseries Cosmic-string calorimetry with a 220\,PeV neutrino}\\[10pt]
Bo-Qiang~Lu\\[2pt]
{\small School of Science, Huzhou Normal University, Huzhou, Zhejiang 313000, P.~R.~China}\\[2pt]
{\small \textsuperscript{*}Electronic address:
\href{mailto:bqlu@huznu.edu.cn}{\textcolor{black}{bqlu@huznu.edu.cn}}}
\end{center}

\vspace{2pt}

\noindent Equation numbers without a prefix refer to the paper. The Appendix
carries the derivations. What follows completes them and compares with earlier
work. The flux that normalizes every estimate is
$E^2\Phi_\nu=5.8\times10^{-8}$~GeV\,cm$^{-2}$\,s$^{-1}$\,sr$^{-1}$ per flavor.

\section*{S1. The injection rate}

The Appendix derives the coefficient of Eq.~\eqref{eq:master}, the redshift
integral and the two conventions for the long-string density. The remaining inputs
are $f_\nu=0.1$, $f_B=10^{-3}$, $\ell=\alpha t$ with $\alpha=0.1$, and
$\Gamma_{\rm GW}=50$, the lower end of the range $50$--$100$ \cite{SM-BOS}. The two
normalizations that follow, Eqs.~\eqref{eq:ratio2} and~\eqref{eq:gammac}, are
checked in Sec.~S2.

The scaling regime fixes how much energy a network releases, not how that energy
is divided among loops, and the flux depends only on the release.

With $\rho_\infty=\zeta\mu H^2$ and $\zeta$ constant, $\rho_\infty\propto t^{-2}$.
In the matter era $H=2/(3t)$, so $\dot\rho_\infty=-2\rho_\infty/t$ and the
continuity equation $\dot\rho_\infty+2H\rho_\infty=-Q$ gives
\begin{equation}
Q=2\rho_\infty\left(\frac{1}{t}-H\right)=\rho_\infty H.
\tag{S1}
\label{eq:qscale}
\end{equation}
In the radiation era $H=1/(2t)$ and the same steps give $Q=2\rho_\infty H$. The
right-hand side contains $\rho_\infty$ and $H$ and nothing else. No length, no
multiplicity, no distribution enters.

The distribution enters only when the rate is turned into a number of bursts. If
the power is released in loops of length $\ell$ and each loop gives up
$\mu\ell$, the burst rate density is $\epsilon_pQ/(\mu\ell)$ and the energy per
burst in one flavor and one logarithmic interval is
$\mu\ell f_\nu/\Delta\ln E$. The product is
$\epsilon_pQf_\nu/\Delta\ln E$, with $\ell$ cancelled, which is
Eq.~\eqref{eq:emissivity}. A distribution that moves the power from small loops to
large ones lowers the burst rate and raises the fluence in the same proportion, and
the diffuse flux is unchanged.

Three quantities do depend on the distribution, and each corrects the efficiency
rather than the rate. The ratio $r$ of Eq.~\eqref{eq:ratio} carries
$\ell/\ell_{\min}$ and decides whether the discharge is efficient at all. The
critical Lorentz factor of Eq.~\eqref{eq:gammac} falls as $1/\ell$ and decides
whether the cusp is beamed. The particle channel must be open at the length that
carries the power. The first two move $\epsilon_p$ and the cone at fixed $Q$. Only
the third can move the flux, and it does so by moving the cut-off.

\section*{S2. The current at the cusp}

Pulsar timing bounds a stochastic background, and the bound is quoted for a network
whose loops radiate only gravitationally. Write $\epsilon_{\rm GW}$ for the share of
the network power that leaves in gravitons, so that the energy density of the
background scales as $G\mu\epsilon_{\rm GW}$ at fixed loop distribution and fixed
spectral shape in the timing band. Then
\begin{equation}
G\mu\left(1-\epsilon_p-\epsilon_V\right)<B,\qquad B=7.9\times10^{-11}
\tag{S2}
\label{eq:timing}
\end{equation}
from the NANOGrav 15~yr data \cite{SM-NANOGrav}, with $\epsilon_V$ the share in
vector radiation, and with $G\mu\epsilon_p=(G\mu)_0$ the floor of Eq.~\eqref{eq:eps}
follows.

Particle emission changes the loop distribution, and the change is mild in the band
that matters. A loop that loses energy to particles lives shorter, so there are more
loops at a given time. The same loop also radiates gravitationally for less of its life. At
fixed frequency the two effects largely cancel, and the main imprint is a cut-off at
high frequency, where the loops no longer reach the length that radiates there
\cite{SM-ASV}. The scaling of Eq.~\eqref{eq:timing} is therefore the form that holds the
distribution fixed, and a full treatment would move the bound by a factor
of order unity rather than by an order of magnitude.

Equation~\eqref{eq:ratio} is the ratio of the particle luminosity of a cusp to its
gravitational luminosity, and the loop length enters through
$\ell/\ell_{\min}$. With $\ell=\alpha t$ and
$\ell_{\min}=\Gamma_{\rm GW}G\mu t$,
\begin{equation}
\frac{\ell}{\ell_{\min}}=\frac{\alpha}{\Gamma_{\rm GW}G\mu}
\tag{S3}
\label{eq:lratio}
\end{equation}
$=4.7\times10^{7}$ at the tension of Eq.~\eqref{eq:fixed}. The coefficient of
Eq.~\eqref{eq:ratio2} then follows with nothing left open. Ref.~\cite{SM-BO09}
gives the ratio of the two luminosities as $50f_{-3}B_{-6}i_c\eta_{10}^{-1}(\ell/\ell_{\min})$, and with
$\eta_{10}=1.22\times10^{9}\sqrt{G\mu}$ and the value above it becomes
$8.2\times10^{-11}f_{-3}B_{-6}i_c/(G\mu)^{3/2}$, which is Eq.~\eqref{eq:ratio2} to the
digit.

The Lorentz factor follows from the same balance. A cusp
boosts the ambient current, $J\simeq0.1e^2B\ell$ at $\gamma=1$ and proportional to
$\gamma$ beyond it, so the boost stops when the current reaches the critical value
$J_{\max}=i_ce\eta$. The Lorentz factor at that point is the ratio of the two,
\begin{equation}
\gamma_c=\frac{i_ce\eta}{0.1e^2B\ell}=\frac{i_c\eta}{0.1eB\alpha t}
=9.5\times10^{5}\,\frac{i_c\sqrt{G\mu}}{B_{-6}}
\tag{S4}
\label{eq:gcfull}
\end{equation}
for $e=0.1$, $\alpha=0.1$ and $t=t_0$. The scaling $\gamma_c\propto\sqrt{G\mu}/B$
carries the physics. A smaller tension means a smaller symmetry-breaking scale, a
smaller maximum current, and a smaller boost before the string discharges.
Evaluated at $\ell_{\min}=\Gamma_{\rm GW}G\mu t$ instead of $\alpha t$, the same
expression scales as $1/\sqrt{G\mu}$, the inverse of Eq.~\eqref{eq:gcfull}, because
$\ell_{\min}\propto G\mu$, and the two differ by $\ell/\ell_{\min}=4.7\times10^{7}$. The
formation length is the one that applies here, since a loop of length $\alpha t$ has
just been produced and carries the current that the field has induced over one
expansion time. The coefficient also carries the product $e\alpha$, which the model
fixes only to a factor of two. Moving it changes where the beaming transition sits
and nothing else, because the efficiency and the tension follow from
Eq.~\eqref{eq:ratio2}, which contains $i_c$ and not $\gamma_c$.

The pair has two consequences. The field that makes the cusp bright also makes
it wide, since $\gamma_c\propto1/B$ while
$r\propto B$, so the cone and the efficiency cannot both be set by the
environment. A loop that forms at $\alpha t$ carries its current at the critical
value to within a factor of six in length, since $\gamma_c=1$
at $\ell\simeq6\alpha t$. A subcritical current requires a string that carries less
charge, not a smaller loop. Figure~2 of the paper shows $\gamma_c$ over the tension for three
fields.

\section*{S3. Inputs and the forecast}

Table~\ref{tab:err} collects the inputs that enter Eq.~\eqref{eq:gmu}. Each enters
as a power, so the budget is multiplicative and a single relation covers the range,
with $N$ the count the event stands for relative to one event:
\begin{equation}
G\mu\epsilon_p=4.3\times10^{-11}
\left(\frac{f_\nu}{0.1}\right)^{-1}
\left(\frac{\zeta}{8.7}\right)^{-1}
\left(\frac{\kappa}{1}\right)^{-1}
\left(\frac{\Delta\ln E}{32}\right)
\left(\frac{N}{1}\right).
\tag{S5}
\label{eq:scale}
\end{equation}

\begingroup
\renewcommand{\thetable}{S\arabic{table}}
\setcounter{table}{0}
\begin{table}[H]
\caption{Inputs of Eq.~\eqref{eq:gmu}, their range, the logarithmic derivative of
$G\mu\epsilon_p$, and the factor the range spans. The reconstructed energy of the
event enters with a vanishing derivative, because $E^2\Phi_\nu$ is flat.}
\label{tab:err}
\footnotesize
\begin{ruledtabular}
\begin{tabular}{lccc}
Input & Range & $\partial\ln(G\mu\epsilon_p)/\partial\ln X$ & Factor \\
\hline
$f_\nu$ & $0.05$--$0.3$ & $-1$ & $2.0$--$0.33$ \\
$\zeta$ & $4$--$20$ & $-1$ & $2.2$--$0.43$ \\
$\kappa$ & $0.6$--$1.4$ & $-1$ & $1.7$--$0.71$ \\
$\Delta\ln E$ & $21$--$32$ & $+1$ & $0.66$--$1$ \\
count $N$ & $0.1$--$3.9$ & $+1$ & $0.11$--$3.9$ \\
$E_{\rm event}$ & $110$--$790$~PeV & $0$ & $1$ \\
\end{tabular}
\end{ruledtabular}
\end{table}
\endgroup

The two dominant entries are the flavor fraction and the count, and they pull in
opposite directions. A larger $f_\nu$ means the same neutrino flux is made with less
power, and a larger count means the event stands for a brighter population. The
vanishing derivative with respect to the reconstructed energy is what makes the
measurement calorimetric. An $E^{-2}$ spectrum carries the
same $E^2\Phi_\nu$ at every energy, so the $220^{+570}_{-110}$~PeV of the event
enters the result only through the assumption that the event lies on the $E^{-2}$
part.

A second event adds an independent draw. For $N$ events the count is known to
$N_{\rm lo}/N=\chi^2_{0.10}(2N)/2N$ and $N_{\rm hi}/N=\chi^2_{0.90}(2N+2)/2N$ at the
same confidence as the single-event range of the work, and each limit maps to a
floor on $\epsilon_p$ through Eq.~\eqref{eq:eps}. The interval is
$0.05$--$0.68$ for one event, $0.19$--$0.52$ for four, $0.25$--$0.46$ for ten,
$0.30$--$0.41$ for thirty and $0.32$--$0.38$ for a hundred, so thirty events narrow
it to $\pm0.06$ and a hundred to $\pm0.03$. At five events a year from the completed
array, thirty events are six years of running.

The forecast assumes that the events are drawn from the same population, so it is a
statement about the network and not about the array alone. A second event at a
different energy would test the $E^{-2}$ shape, and one at a much lower energy would
test the cut-off.

An isotropic discharge has no beaming penalty and no beaming gain. The number of
bursts that point at us and the fluence per burst move in opposite directions, and
their product is the diffuse flux. What changes is the burst rate a detector can
resolve, so a search for isolated $10^{17}$--$10^{18}$~eV events is sensitive to
that rate and not to the flux.

The cascade is assumed to put equal energy in every logarithmic interval, which is
what an $E^{-2}$ spectrum does. For $dN/dE\propto E^{-2-\delta}$ the energy per
interval at the event, $E=2.2\times10^{8}$~GeV, is smaller by
\begin{equation}
F(\delta)=\frac{\delta\,\Delta\ln E}
{\exp[\delta\ln(E/1~{\rm GeV})]-\exp[-\delta\ln(E_{\max}/E)]},
\tag{S6}
\label{eq:index}
\end{equation}
which is $0.93$ at $\delta=-0.1$ and $0.49$ at $\delta=+0.1$, so the tension rises
to $4.6$ or $8.8\times10^{-11}$ and the floor to $0.37$ or $0.53$. At
$\delta=\pm0.2$ the floor spans $0.52$ to $0.80$. The calorimetric argument
survives an index two tenths from two, at the price of an efficiency closer to
unity.

\section*{S4. Comparison with earlier calculations}

Refs.~\cite{SM-BO09,SM-Lunardini,SM-ASV} work in the burst framework. A loop
production function is integrated over length and time, each loop is weighted by a
burst rate, a beaming factor and an ambient field, and the bursts are summed. The
flux then depends on several powers of $G\mu$ and on quantities that are
modeled rather than measured.

The numerical comparison with Ref.~\cite{SM-BO09} is the sharpest, because its
result does not depend on $\eta$. Summed over flavors, its flux is
$E^2J_\nu=6.6\times10^{-8}i_cB_{-6}f_{-3}$, or $2.2\times10^{-8}$ per flavor at
unit $i_cB_{-6}f_{-3}$, a factor of $2.6$ below the KM3NeT flux. The logarithmic
factor estimated there as $30$ is the $\Delta\ln E$ of Eq.~\eqref{eq:master}, which
Eq.~\eqref{eq:width} sharpens to $32$. Setting
$4.3\times10^{5}G\mu f_\nu/\Delta\ln E$ with $f_\nu=0.1$ and $\Delta\ln E=32$
equal to $2.2\times10^{-8}$ gives $G\mu=1.6\times10^{-11}$, where
Eq.~\eqref{eq:gammac} gives $\gamma_c=3.8$, the crossing quoted in the work. Below
that tension the two part company. At $\eta=10^{10}$~GeV, where
$G\mu=6.7\times10^{-19}$, the burst calculation still returns $6.6\times10^{-8}$
while the budget allows $1.2\times10^{-15}$, eight orders less, because that flux
carries no $\gamma_c$ and Eq.~\eqref{eq:master} falls in proportion to $G\mu$.

Ref.~\cite{SM-Lunardini} radiates moduli from kinks rather than the current from
cusps, and the difference shows in the scaling. The flux is then
$E^2J_\nu\simeq1.7\times10^{-7}\mu_{-17}^{3/7}\bar\alpha^2E_{11}^{1/7}/[p(\Gamma/50)^{11/7}]$,
a $3/7$ power of the tension rather than the first power of
Eq.~\eqref{eq:master}, with a normalization set by the modulus coupling $\bar\alpha$
and its mass. The flux changes by a factor of fifty when
the tension changes by four orders of magnitude, so a measurement of the flux
constrains a combination of $\bar\alpha$, $m$ and $G\mu$ and not the tension
itself.
Its spectrum also starts at $E\gtrsim10^{11}$~GeV, three orders of magnitude above
the $2.2\times10^{8}$~GeV of the event, because the kink radiation is beamed into a
fan of solid angle $2\pi(kL)^{-1/3}$ at a Lorentz factor of order $10^{13}$.

Ref.~\cite{SM-ASV} asks what particle emission does to the loop distribution, and
finds that the number density stops scaling and a high-frequency cut-off appears in
the gravitational-wave background, at a frequency above the band of current and
planned detectors. Kinks and cusps behave alike, and under mild assumptions about the particle
physics the diffuse gamma-ray background does not constrain $G\mu$.
Particle emission does not remove the loops that source the timing band, so
Eq.~\eqref{eq:timing} applies and the gravitational channel can be bounded while
the neutrino flux measures the particle one.

What none of the three does is use the scaling of $\rho_\infty$ to fix the injection
rate, and that is what makes Eq.~\eqref{eq:master} independent of the loop
distribution.

\section*{S5. Other channels}

Table~\ref{tab:mech} collects the mechanisms proposed to power the flux. Cusp
annihilation \cite{SM-HSW87,SM-VHS} releases the bulk energy of the loop with a
fragmentation spectrum close to $E^{-2}$ below its peak, and it alone reaches
order-unity efficiency without a new parameter. A fireball is excluded even
after its own Lorentz boost, which lifts a GeV plasma into the TeV range and
leaves the flux at $220$~PeV short by $e^{-2.8\times10^{4}}$
\cite{SM-GV16}, and the remaining three replace the tension by a free parameter
and lose the prediction in the process.

\begingroup
\renewcommand{\thetable}{S\arabic{table}}
\setcounter{table}{1}
\setlength{\intextsep}{2pt plus 1pt}
\begin{table}[H]
\caption{Mechanisms proposed to power the flux at $220$~PeV, the efficiency each
allows, the parameter that replaces $G\mu$, and the value the flux requires. The two
cusp rows are fixed by $G\mu\epsilon_p=(2.8$--$4.3)\times10^{-11}$. In the last three
the flux fixes a parameter that the model does not predict. References are in the
work.}
\label{tab:mech}
\footnotesize
\begin{ruledtabular}
\begin{tabular}{llcc}
Mechanism & $\epsilon_p$ & Parameter & Value required \\
\hline
Current-carrying cusp & $r/(1+r)\to1$ & $G\mu$ & $(2.8$--$4.3)\times10^{-11}$ \\
Cusp annihilation & $0.1$--$1$ & $G\mu$ & $(2.8$--$43)\times10^{-11}$ \\
Fireball & --- & --- & excluded, $e^{-2.8\times10^{4}}$ \\
Vorton & free & $\Omega_v/\tau$ & $3.2\times10^{-27}$~s$^{-1}$ \\
Necklace & $\sim1$ & monopole mass & $E_{\max}$ up to the GUT scale \\
Cusp $\to$ PBH & --- & $M/\mu$ & $5$~GeV$^{-1}$ \\
\end{tabular}
\end{ruledtabular}
\end{table}
\endgroup

A current-carrying loop can be stabilized by its angular momentum and become a
vorton \cite{SM-Carter,SM-MartinsShellard,SM-Lemperiere}. Vortons form with an abundance $\Omega_v$ set by the formation efficiency rather than by $G\mu$, and decay with an independent lifetime $\tau$, so Eq.~\eqref{eq:master} is replaced by
\begin{equation}
E^2\Phi_\nu=5.5\times10^{21}\;\frac{\Omega_v}{\tau}\;
\frac{f_\nu}{\Delta\ln E},
\tag{S7}
\label{eq:vorton}
\end{equation}
with $\tau$ in seconds. Reproducing $5.8\times10^{-8}$ requires
$\Omega_v/\tau\simeq3.2\times10^{-27}$~s$^{-1}$, or $\tau\simeq3.1\times10^{25}$~s
for $\Omega_v=0.1$, some seventy million Hubble times. The fit reproduces the
flux at the cost of the prediction. A necklace \cite{SM-BV97} behaves in the same way
with the monopole mass in place of the lifetime, and $E_{\max}$ can be placed
anywhere between the PeV and the grand-unified scale, which leaves necklaces the most
flexible of the non-standard options \cite{SM-BV97,SM-Leblond,SM-Matsuda}.

A hole of mass $M$ has $T_{\rm BH}\simeq1.1$~TeV\,$(10^{10}\,{\rm g}/M)$ and
lives for $\tau\simeq8\times10^{4}$~s\,$(M/10^{10}\,{\rm g})^3$. Emission at
$220$~PeV requires $M\simeq5\times10^{4}$~g and $\tau\simeq10^{-11}$~s, and the
lifetime is not the obstacle, since cusps collapse throughout the history of the
network. The mass is the obstacle. A loop that carries the network power has
$\ell=\alpha t\simeq4\times10^{40}$~GeV$^{-1}$, so the hole would have to contain
$M/\mu\simeq5$~GeV$^{-1}$ of string, some $10^{14}$ times the width of the
string and thirty-five orders below the overlap $\sqrt{G\mu}\,\ell$. A region of
either scale misses by those factors, the width collapsing to $10^{-10}$~g and the
overlap to $10^{39}$~g.

\end{document}